\documentclass{llncs}
\usepackage{url}
\usepackage{multirow}
\usepackage{balance}
\usepackage[figuresright]{rotating}
\usepackage{booktabs}
\usepackage{hyphsubst}
\usepackage{graphicx}
\usepackage{amsmath}
\graphicspath{{images/}}
\usepackage{hhline}
\usepackage[caption = false]{subfig}

\usepackage{url}
\usepackage{rotating}
\usepackage{amsmath}
\usepackage{cancel}
\usepackage{amssymb}
\usepackage{enumitem}
\usepackage{pifont}

\usepackage{multirow}
\usepackage{array}
\usepackage{placeins}

\usepackage{rotating}
\usepackage{booktabs}
\usepackage{rotating}
\usepackage{hhline}
\usepackage{hyphsubst}
\usepackage{soul}

\newlength{\Oldarrayrulewidth}

\begin{document}

\title{Utilizing Online Social Network and Location-Based Data to Recommend Products and Categories in Online Marketplaces}

\author{Emanuel Lacic\inst{1} \and Dominik Kowald\inst{1} \and Lukas Eberhard\inst{2} \and Christoph Trattner\inst{1} \and Denis Parra\inst{3} \and Leandro Marinho\inst{4}}
\institute{
	Know-Center, 
	Graz University of Technology, Graz, Austria 
	\\ \email{\{elacic,dkowald,ctrattner\}@know-center.at} 
	\and
	IICM,  
	Graz University of Technology, Graz, Austria
	\\ \email{lukas.eberhard@tugraz.at} 
	\and 
	Pontificia Universidad Católica de Chile, 
	Santiago, Chile
	\\\email{dparra@ing.puc.cl}
	\and 
	UFCG, 
	Campina Grande, Brasil
	\\\email{lbmarinho@dsc.ufcg.edu.br}
	}

\maketitle

\begin{abstract}
Recent research has unveiled the importance of online social networks for improving the quality of recommender systems and encouraged the research community to investigate better ways of exploiting the social information for recommendations. 
To contribute to this sparse field of research, in this paper we exploit users' interactions along three data sources (marketplace, social network and location-based) to assess their performance in a barely studied domain: recommending products and domains of interests (i.e., product categories) to people in an online marketplace environment. To that end we defined sets of content- and network-based user similarity features for each data source and studied them isolated using an user-based Collaborative Filtering (CF) approach and in combination via a hybrid recommender algorithm, to assess which one provides the best recommendation performance. Interestingly, in our experiments conducted on a rich dataset collected from SecondLife, a popular online virtual world, we found that recommenders relying on user similarity features obtained from the social network data clearly yielded the best results in terms of accuracy in case of predicting products, whereas the features obtained from the marketplace and location-based data sources also obtained very good results in case of predicting categories. This finding indicates that all three types of data sources are important and should be taken into account depending on the level of specialization of the recommendation task.

\end{abstract}
\newpage

\section{Introduction} \label{sec:introduction}

Research on recommender systems has gained tremendous popularity in recent years. Especially since the hype of the social Web and the rise of social media and networking platforms such as Twitter or Facebook, recommender systems are acknowledged as an essential feature helping users to, for instance, discover new connections between people or resources. Especially in the field of e-commerce sites, i.e., online marketplaces, current research is dealing with the improvement of the prediction task in order to recommend products that are more likely to match peoples preferences.

Typically, these online systems calculate personalized recommendations using only one data source, namely marketplace data (e.g., implicit user feedback such as previously viewed or purchased products - see also e.g., \cite{Zhang2013,GuoLeskovec2011,trattner2014}). Although this approach has been well established and performs reasonably well, nowadays, online marketplaces often also have the opportunity to leverage additional information about the users coming from social and location-based data sources (e.g., via Facebook-connect). Even though previous research has shown that this kind of data can be useful in the wide field of recommender systems (see Section \ref{sec:relatedwork}), it remains an open problem how to fully exploit these additional data sources (social network and location-based data) to improve the recommendation task in online marketplaces. 


Moreover, it is often not the most important thing in online marketplaces to predict the exactly right products to the users but to suggest domains of interests (i.e., product categories) the users could like and could use for further browsing (e.g, \cite{ma2011,Zhang2013}). Thus, it is not only important to investigate to which extend social and location-based data sources can be used to improve product recommendations but also to which extend this data can also be used for the recommendation of product categories.

To contribute to this sparse field of research, in this paper we present a first take on this problem in form of a research project that aims at understanding how different sources of interaction data can help in recommending products and categories to people in an online marketplace. In this respect, we are  particularly interested in studying the efficiency of different user similarity features derived from various dimensions, not only from the marketplace but also from online social networks and location-based data to recommend products and categories to people via a user-based Collaborative Filtering (CF) approach (we have chosen a user-based CF approach since user-based CF is not only a well-established recommender algorithm but also allows us to incorporate various user-based similarity features coming from different data sources, which has been shown to play an important role in making more accurate predictions \cite{Jamali2010Trust,ma2011}). Specifically, we raise the following two research questions:

\begin{itemize}
	\item \textit{RQ1}: To which extent can user similarity features derived from marketplace, social network and location-based data sources be utilized for the recommendation of products and categories in online marketplaces? \\
	\item \textit{RQ2}: Can the different marketplace, social network and location-based user similarity features and data sources be combined in order to create a hybrid recommender that provides more robust recommendations in terms of prediction accuracy, diversity and user coverage?
\end{itemize}

In order to address these research questions, we examined content-based and network-based user similarity feature sets for user-based CF over three data sources (social, transactional, and location-based data) as well as their combinations using a hybrid recommender algorithm and assessed the results via a more comprehensive set of recommender evaluation metrics than previous works. The study was conducted using a large-scale dataset crawled from the virtual world of SecondLife. In this way, we could study the utility of each user similarity feature separately as well as combine them in the form of hybrid approaches to show which combinations, per data source and globally, provide the best recommendations in terms of recommendation accuracy, diversity and user coverage. Summing this up, the contributions of this work are the following:

\begin{itemize}
	\item Contrary to previous work in this area \cite{Zhang2013,GuoLeskovec2011}, we not only employ one source of data (marketplace transactions) for the problem of predicting product purchases but show how data coming from three different sources (marketplace, social and location-based data) can be exploited in this context. \\
	\item In contrast to related work in the field, we provide also an extensive evaluation of various content-based and network-based user similarity features via user-based Collaborative Filtering as well as their combinations via a hybrid recommender approach.  \\
	\item Finally, we also provide evidence to what extent top-level and sub-level purchase categories can be predicted which is in contrast to previous work (e.g., \cite{Zhang2013}) where the authors only focused on the problem recommending top-level categories to the users.
\end{itemize}
To the best of our knowledge, this is the first study that offers such a comprehensive user similarity feature selection and evaluation for product and category recommendation in online marketplaces.

Overall, the paper is structured as follows: we begin by discussing related work in Section \ref{sec:relatedwork}. Then we present the datasets (Section \ref{sec:datasets}) and the feature description (Section \ref{sec:features}) used in our extensive evaluation. After that, we present our experimental setup in Section \ref{sec:expsetup} and show the results of our experiments in Section \ref{sec:eval}. Finally, on Section \ref{sec:con} we conclude the paper and discuss the outlook.

\section{Related Work}
\label{sec:relatedwork}

Most of the literature that leverages social data for recommendations is focused on recommending users, (e.g., \cite{GuoLeskovec2011,bischoff2012we}), tags (e.g., \cite{Feng2012}) or points-of-interest (e.g, \cite{ma2011}), although some works have exploited social information for item or product recommendation, being the most important ones model-based. Jamali et al. \cite{Jamali2010Trust} introduced SocialMF, a matrix factorization model that incorporates social relations into a rating prediction task, decreasing RMSE with respect to previous work. Similarly, Ma et al. \cite{ma2011} incorporated social information in two models of matrix factorization with social regularization, with improvements in both MAE and RMSE for rating prediction. Among their evaluations, they concluded that choosing the right similarity feature between users plays an important role in making a more accurate prediction.

On a more general approach, Karatzoglou et al. \cite{secofi2013} use implicit feedback and social graph data to recommend places and items, evaluating with a ranking task and reporting significant improvements over past related methods. Compared to these state-of-the-art approaches, our focus on this paper is at providing a richer analysis of feature selection (similarity features) with a more comprehensive evaluation than previous works, and in a rarely investigates domain: product recommendation in a social online marketplace. For instance, in Guo et al. \cite{GuoLeskovec2011} or Trattner et al. \cite{trattner2014} the authors leveraged social interactions between sellers and buyers in order to predict sellers to customers. Other relevant work in this context is the study of Zhang \& Pennacchiotti \cite{Zhang2013} who showed how top-level categories can be better predicted in a cold-start setting on eBay by exploiting the user's ``likes'' from Facebook. 

\section{Datasets}
\label{sec:datasets}
In our study we relay on three datasets\footnote{\textbf{Note:} The datasets could be obtained by contacting the fourth author of this work.} obtained from the virtual world SecondLife (SL). The main reason for choosing SL over real world sources are manifold but mainly due to the fact that currently there are no other datasets available that comprise marketplace, social and location data of users at the same time. For our study we focused on users who are contained in all three sources of data, which are 7,029 users in total. To collect the data (see Table \ref{tab:slstats}) we crawled the SL platform as described in our previous work \cite{lacic2014,steurer2013}.
\begin{table}[t!]
\small
\setlength{\tabcolsep}{3pt}
\centering
\begin{tabular}{ll}
\specialrule{.2em}{.1em}{.1em}
Marketplace Dataset (Market) & \\\hline
Number of products                 & $30,185$                        \\ 
Number of products with categories  &  $24,276$ \\

Number of purchases                & $39,055$                     \\
Number of purchases with categories  &  $31,164$ \\

Mean number of purchases per user                & $5.56$                     \\
Mean number of purchases per products                 &  $1.29$                        \\  
Mean number of categories per product & $2.86$ \\
Number of top-level categories & $23$ \\
Number of low-level categories & $532$ \\

Mean number of top-level categories purchase & $1,354.96$ \\
Mean number of low-level categories purchase & $58.58$ \\

Number of sellers & $8,149$ \\
Mean number of purchases per seller & $3.70$ \\
\specialrule{.2em}{.1em}{.1em}
Online Social Network Dataset (Social)                 &     \\ \hline
Number of interactions			& $490,236$ \\
Mean number of interactions                & $69.75$ \\
Number of groups                 & $39,180$                         \\
Mean number of groups per user               & $9,419$                          \\
Number of interests                & $5.57$                         \\ 
Mean number of interests per user               & $1.34$ \\ 
\specialrule{.2em}{.1em}{.1em}
Location-based Dataset (Location)                 &     \\ \hline

Number of different favorite locations & $10,538$ \\
Mean number of favorite locations per user & $5.77$ \\

Number of different shared locations& $5,736$ \\
Mean number of shared locations per user & $1.94$ \\

Number of different monitored locations & $1,887$ \\
Mean number of monitored locations per user & $6.52$ \\

\specialrule{.2em}{.1em}{.1em}
\end{tabular}
\vspace{2mm}
\caption{Basic statistics of the SL datasets used in our study.}
\vspace{4mm}
\label{tab:slstats}
\end{table}

\begin{figure*}[ht!]
  \centering
  		 \subfloat[SecondLife store]{ 
				\includegraphics[width=0.49\textwidth,height=0.49\textwidth]{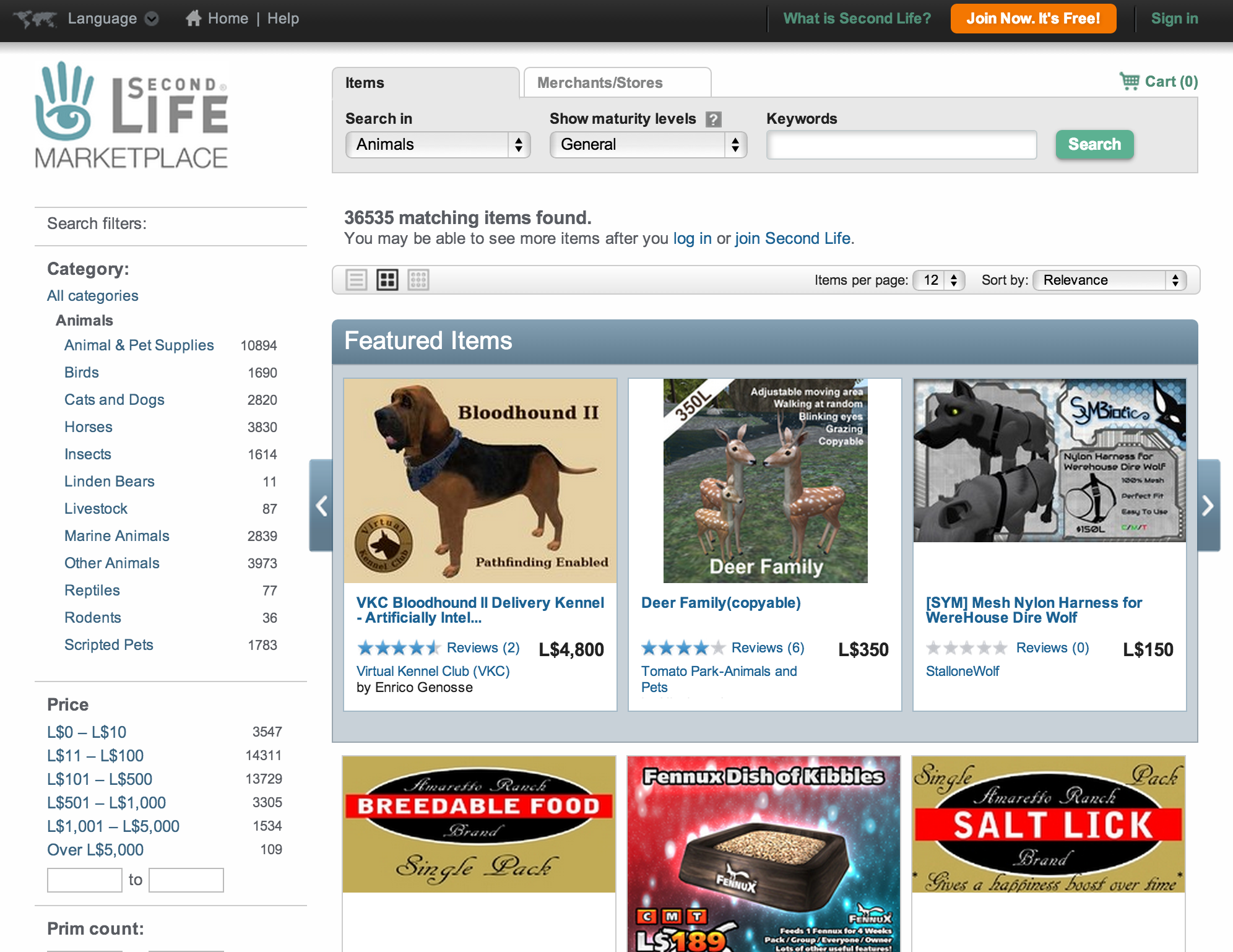}%
		 }
		 \subfloat[SecondLife social stream]{ 
				\includegraphics[width=0.49\textwidth,height=0.49\textwidth]{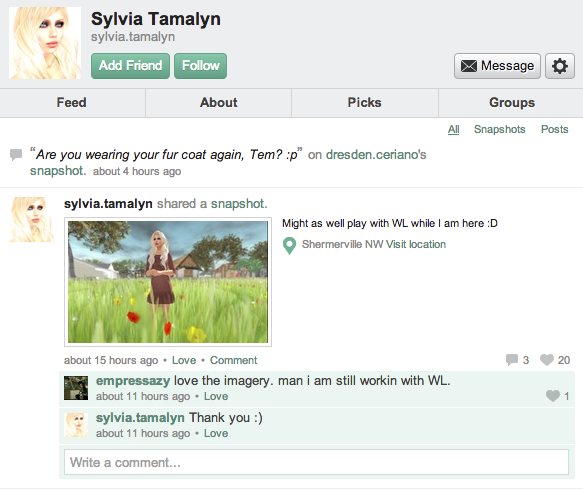}%
		 }
     \caption{Examples for a store in the marketplace and a social stream of an user in the online social network of the virtual world SecondLife.}
    \label{fig:secondlife}
\end{figure*}

\subsection{Marketplace Dataset}

Similar to eBay, SecondLife provides an online trading platform where users can trade virtual goods with each other. Every seller in the SL marketplace\footnote{\url{https://marketplace.secondlife.com/}} owns her own seller's store and publicly offers all of the store's products classified into different categories (a hierarchy with up to a maximum of four different categories per product). Furthermore, sellers can apply meta-data such as price, title or description to their products. Customers in turn, are able to provide reviews to products.

We extracted 29,802 complete store profiles, with corresponding 39,055 trading interactions, and 2,162,466 products, out of which 30,185 were purchased. From the purchased products, 24,276 are described using categories which are differentiated in top-level categories and low-level categories (i.e, the assigned product categories on the lowest possible level of the category hierarchy). 
An example of a product in the marketplace of SL is shown in the first image of Figure \ref{fig:secondlife}.

\subsection{Online Social Network Dataset}

The online social network MySecondLife\footnote{\url{https://my.secondlife.com/}} is similar to Facebook with regard to postings: users can interact with each other by sharing text messages and commenting or loving (= liking) these messages. From the extracted 7,029 complete user profiles, we gathered 39,180 different groups users belong to, 9,419 different interests users defined for themselves and 490,236 interactions between them. The second image of Figure \ref{fig:secondlife} shows an example of an user profile in the online social network of SL.

\subsection{Location-based Dataset}

The world of Second Life is contained within regions, i.e., locations which are independent from each other. Overall, we extracted three different sources of location-based data in our experiments:

\begin{enumerate}[label=\alph*)]
\item Favored Locations: Every user of SL can specify up to 10 so-called ``Picks'' in their profile representing her favorite locations that other users can view in the user's MySecondLife profile. We found that the extracted users picked 40,558 locations from 10,538 unique locations;\\
\item Shared Locations: Users in SL can also share information about their current in-world position through in-world pictures called "snapshots", which also include in-world GPS information (similar as Foursquare). Overall, we identified 13,637 snapshots in 5,736 unique locations;\\
\item Monitored Locations: As in real life, users in SL can create events in the virtual world and publicly announce them in a public event calendar. We collected these events, with an accurate location and start time, and extracted 157,765 user-location-time triples, with 1,887 unique locations.
\end{enumerate}

\section{Feature Description} 
\label{sec:features}
As shown in our previous work (e.g., \cite{steurer2013}), similarities between users can be derived in two different ways: either we calculate similarities between users on the content (= meta-data) provided directly by the user profiles or on the network structure of the user profiles interacting with each other. In the following sections we present more details about these ideas.

\subsection{Content-based User Similarity Features}

We define our set of content-based user similarity features based on different types of entities or meta-data information that are directly associated with the user profiles in our data sources. In the case of the marketplace dataset these entities are purchased products, product categories and sellers of the products, in the case of the social network these are groups and interests the users have assigned, and in the case of the location-based data source these are favored locations, shared locations and monitored locations. Formally, we define the entities of a user $u$ as $\Delta(u)$ in order to calculate the similarity between two users, $u$ and $v$.

The first content-based user similarity feature we induce is based on the entities two users have in common. It is called \emph{Common Entities} and is given by: 
  \begin{align}
  sim(u, v) = |\Delta(u) \cap \Delta(v)|
  \end{align}
The second similarity feature, \emph{Total Entities}, is defined as the union of two users' entities and is calculated by:
  \begin{align}
  sim(u, v) = |\Delta(u) \cup \Delta(v)|
  \end{align}
These two user similarity features are combined by \emph{Jaccard's Coefficient for Entities} as the number of common entities divided by the total number of entities: 
  \begin{align}
sim(u, v) = \frac{|\Delta(u) \cap \Delta(v)|}{|\Delta(u) \cup \Delta(v)|}
  \end{align}
\subsection{Network-based User Similarity Features}

In our experiments we consider all networks as an undirected graph $G \langle V, E \rangle$ with $V$ representing the user profiles and $e = (u, v) \in E$ if user $u$ performed an action on $v$ (see also \cite{steurer2013}). In the case of the social network, these actions are defined as social interactions, which are a combination of likes, comments and wallposts. In the case of the location-based dataset, actions between users are determined if they have met each other in the virtual world at the same time in the same location\footnote{\textbf{Note:} We derived the networks in our study from the location-based dataset only for the monitored locations, since the exact timestamps are not available for the favored nor the shared locations in the datasets.}. Furthermore, the weight of an edge $w_{action}(u, v)$ gives the frequency of a specific action between two users $u$ and $v$. Finally, this network structure also let us determine the neighbors of users in order to calculate similarities based on this information. We define the set of neighbors of a node $v \in G$ as $\Gamma(v)$ = $\{u~|~(u, v) \in E \}$.

The first network-based user similarity feature we introduce, uses the number of \emph{Directed Interactions} between two users and is given by:
 \begin{align}
 sim(u, v) = w_{action}(u, v)
 \end{align}
 In contrast to \emph{Directed Interactions}, the following user similarity features are based on the neighborhood of two users: The first neighborhood similarity feature is called \emph{Common Neighbors} and represents the number of neighbors two users have in common: 
\begin{align}
sim(u, v) = |\Gamma(u) \cap \Gamma(v)|
  \end{align}
To also take into account the total number of neighbors of the users, we introduced \textit{Jaccard's Coefficient for Common Neighbors}. It is defined as:
\begin{align}
sim(u,v) = \frac{|\Gamma(u) \cap \Gamma(v)|}{|\Gamma(u) \cup \Gamma(v)|}
  \end{align}
A refinement of this feature was proposed as \emph{Adamic Adar} \cite{adamic2003friends}, which adds weights to the links since not all neighbors in a network have the same tie strength:
\begin{align}
sim(u, v) = \sum\limits_{z \in \Gamma(u) \cap \Gamma(v)} \frac{1}{log(|\Gamma(z)|)}
  \end{align}
Another related similarity feature introduced by Cranshaw et al. \cite{cranshaw2010bridging}, called \textit{Neighborhood Overlap}, measures the structural overlap of two users. Formally, this is written as: 
\begin{align}
sim(u,v) = \frac{|\Gamma(u) \cap \Gamma(v)|}{|\Gamma(u)| + |\Gamma(v)|}
  \end{align}
 The \emph{Preferential Attachment Score}, first mentioned by Barabasi et al. \cite{barabasi1999emergence}, is another network-based similarity feature with the goal to prefer active users in the network. This score is the product of the number of neighbors of each user and is calculated by:
	\begin{align}
	sim(u, v) = |\Gamma(u)| \cdot |\Gamma(v)|
  \end{align}

\section{Experimental Setup}
\label{sec:expsetup}
In this section we provide a detailed description of our experimental setup. First, we describe the recommender approaches we have chosen in order to evaluate our three data sources (marketplace, social network and location-based data) as well as our derived user similarity features for the task of recommending products and categories. Afterwards, we describe the evaluation methodology and the metrics used in our study.

\subsection{Recommender Approaches}
\label{sec:algos}

In this subsection we describe the recommender approaches we have used to tackle our research questions described in the introductory section of this paper. All mentioned approaches have been implemented into our scalable big data social recommender framework \textit{SocRec} \cite{lacic2014}, an open-source framework which can be obtained from our Github repository\footnote{\url{https://github.com/learning-layers/SocRec}}.

\subsubsection{Baseline.} As baseline for our study, we used a simple MostPopular approach recommending the most popular products or categories in terms of purchase frequency to the users.

\subsubsection{Recommending Products.} The main approach we adopt to evaluate our data sources and user similarity features for the task of recommending products is a \textit{User-based Collaborative Filtering (CF)} approach. The basic idea of this approach is that users who are more similar to each other, e.g., have similar taste, will likely agree or rate on other resources in a similar manner \cite{schafer2007collaborative}. Out of the different CF approaches, we used the non-probabilistic user-based nearest neighbor algorithm where we first find the $k$-nearest similar users and afterwards recommend the resources of those user as a ranked list of top-$N$ products to the target user that are new to her (i.e., she has not purchased those products in the past). As outlined before, we have chosen this approach since user-based CF is not only a well-established recommender algorithm but also allows us to incorporate various user-based similarity features coming from different data sources, which has been shown to play an important role in making more accurate predictions \cite{Jamali2010Trust,ma2011}.

The similarity values of the user pairs $sim(u, v)$ are calculated based on the user similarity features proposed in Section \ref{sec:features} (i.e., constructing the neighborhood). Based on these similarity values, each item $i$ of the $k$ most similar users for the target user $u$ is ranked using the following formula \cite{schafer2007collaborative}:
\begin{align}
		pred(u, i) = \sum\limits_{v \in neighbors(u)}{sim(u, v)}
\end{align}

\subsubsection{Recommending Categories.} For the task of recommending categories and in contrast to previous work (e.g., \cite{Zhang2013}), we are not only focusing here on the prediction of top-level categories but also on the prediction of low-level categories. The prediction of categories was implemented as an extension of product predictions. Thus, for each product in the list of recommended products (i.e., the products obtained from the $k$-nearest neighbors of user $u$ based on a user similarity feature), we extracted the assigned category on the highest level in the case of predicting top-level categories and the assigned category on the lowest level in the case of predicting low-level categories. Afterwards, we assigned a score to each extracted category $e_{i}$ in the set of all extracted categories $E_u$ for the target user $u$ based on a similar method as proposed in \cite{Zhang2013}:

\begin{align}
		pred(u, e_{i}) = \frac{ purc(E_u, e_{i}) } { \sum\limits_{ e \in E_u }{purc(E_u, e)}}
\end{align}
where $purc(E_u, e_{i})$ gives the number of times the category $e_{i}$ occurs in the set of all extracted categories $E_u$ for user $u$.

\subsubsection{Combining User Similarity Features and Data Sources.} To further explore how to combine our data sources and features for recommendation, we investigated different hybridization methods (see also \cite{bostandjiev2012tasteweights}). 
The hybrid approach chosen in the end is known as \textit{Weighted Sum}. The score of each recommended item  in the \textit{Weighted Sum} algorithm is calculated as the weighted sum of the scores for all recommender approaches. 
It is given by:
  \begin{align}
		W_{rec_{i}} = \sum\limits_{s_{j} \in S}{(W_{rec_{i}, s_{j}} \cdot W_{s_{j}})}
  \end{align}
where the combined weighting of the recommended item $i$, $W_{rec_{i}}$, is given by the sum of all single weightings for each recommended item in an approach $W_{rec_{i}, s_{j}}$ multiplied by the weightings of the recommender approaches $W_{s_{j}}$. We weighted each recommender approach $W_{s_{j}}$ based on the nDCG@10 value obtained from the individual approaches. 


We also experimented with other hybrid approaches, known as \textit{Cross-source} and \textit{Mixed Hybrid} \cite{bostandjiev2012tasteweights}. 
However, these approaches have not yielded better results than the \textit{Weighted Sum} algorithm.

\subsection{Evaluation Method and Metrics}
\label{sec:metrics}

To evaluate the performance of each approach in a recommender setting, we performed a number of off-line experiments. Therefore, we split the SL dataset in two different sets (training and test set) using a method similar to the one described in \cite{lacic2014}, i.e., for each user we withheld 10 purchased products from the training set and added them to the test set to be predicted. Since we did not use a $p$-core pruning technique to prevent a biased evaluation, there are also users with less than 10 relevant products. We did not include these users into our evaluation since they did not have enough purchase data available that could be used to produce reasonable recommendations based on the marketplace data, although this data is worthwhile for our user-based CF approach to find suitable neighbors. Thus, we used a post-filtering method, where all the recommendations were still calculated on the whole datasets but accuracy estimates were calculated only based on these filtered user profiles (= 959 users).


To finally quantify the performance of each of our recommender approaches, we used a diverse set of well-established metrics in recommender systems \cite{SmythMcClave01,herlocker2004evaluating}. These metrics are as follows: 

\textbf{Recall (R@k)} is calculated as the number of correctly recommended products divided by the number of relevant products, where $r_u^k$ denotes the top $k$ recommended products and $R_u$ the list of relevant products of a user $u$ in the set of all users $U$. Recall is given by \cite{van1974foundation}:  
   \begin{align}
		R@k = \frac{ 1 }{ |U| } \sum\limits_{ u \in U }{ (\frac{| r_u^k \cap R_{u} |}{ |R_{u}| }) }
  \end{align}
			
\textbf{Precision (P@k)} is calculated as the number of correctly recommended products divided by the number of recommended products $k$. Precision is defined as \cite{van1974foundation}:
  \begin{align}
		P@k = \frac{ 1 }{ |U| } \sum\limits_{ u \in U }{ (\frac{| r_u^k \cap R_{u} |}{ k }) }
  \end{align}

\textbf{Normalized Discounted Cumulative Gain (nDCG@k)} is a ranking-dependent metric that not only measures how many products can be correctly predicted but also takes the position of the products in the recommended list with length $k$ into account. The nDCG metric is based on the \textit{Discounted Cummulative Gain (DCG@$k$)} which is given by \cite{ParraSahebi}: 
   \begin{align}
      DCG@k = \sum\limits_{k = 1}\limits^{|r_u^k|} (\frac{2 ^ {B(k)} - 1}{log_{2}(1 + k)})
   \end{align}
where $B(k)$ is a function that returns 1 if the recommended product at position $i$ in the recommended list is relevant. nDCG@$k$ is calculated as DCG@$k$ divided by the ideal DCG value iDCG@$k$ which is the highest possible DCG value that can be achieved if all the relevant products would be recommended in the correct order. Taken together, it is given by the following formula \cite{ParraSahebi}:
  \begin{align}
      nDCG@k = \frac{ 1 }{ |U| } \sum\limits_{ u \in U }{ (\frac{DCG@k}{iDCG@k}) }
\end{align}
 
\textbf{Diversity (D@k)}, as defined in \cite{SmythMcClave01}, can be calculated as the average dissimilarity of all pairs of resources in the list of recommended products $r_u^k$. Given a distance function $d(i,j)$ that is the distance, or the dissimilarity between two products $i$ and $j$, $D$ is given as the average dissimilarity of all pairs of products in the list of recommended products \cite{SmythMcClave01}:
  
  \begin{align}
		D@k = \frac{ 1 }{ |U| } \sum\limits_{ u \in U }{(\frac{ 1 }{ k \cdot (k - 1) } \sum\limits_{ i \in R }{ \sum\limits_{ j \in r_u^k, j \neq i  }{  }{ d(i,j) } })}
	\end{align}
	
\textbf{User Coverage (UC)} is defined as the number of users for whom at least one product recommendation could have been calculated ($|U_r|$) divided by total number of users $|U|$ \cite{ge2010beyond}:
\begin{align}
	UC = \frac{|U_r|}{|U|}
\end{align}

All mentioned performance metrics are calculated and reported based on the top-10 recommended products.

\begin{sidewaystable*}
\begin{center}
\scalebox{0.8}{
\begin{tabular}{c|lllll|lll|lll|ll}
\specialrule{.2em}{.1em}{.1em}
		\multicolumn{3}{c}{\centering{}}  & \multicolumn{3}{c}{\centering{Products}} & \multicolumn{3}{c}{\centering{low-level categories}} & \multicolumn{3}{c}{\centering{top-level categories}} & &\\ \hline

  \multicolumn{2}{c}{} & User Sim. Feature &  		$nDCG@10$ 			   & $P@10$ 		& $R@10$  & $nDCG@10$ 			   & $P@10$ 		& $R@10$ & $nDCG@10$ 			   & $P@10$ 		& $R@10$ & $D@10$ 	& $UC$ \\\hline
 \multicolumn{3}{c}{Most Popular} & $.0048$  & $.0037$ & $.0047$  & $.0185$  & $.0207$ & $.0157$ & $.2380$ & $.2730$ & $.2221$ & $.6392$ & $100.0\%$\\
  \hline\hline
  \multirow{8}{15pt}{\centering{\begin{sideways}\centering{Market}\end{sideways}}} &  
  \multirow{8}{15pt}{\centering{\begin{sideways}\centering{Content}\end{sideways}}} 
  &  Common Purchases 							& $.0097$ & $.0073$ & $.0094$ & $.0724$ & $.0641$ & $.0757$ & $.4557$ & $.3884$ & $.4636$ & $.5892$ & $90.51\%$ \\
  && Common Sellers  								& $.0146$ & $.0102$ & $.0142$ & $.1119$ & $.1005$ & $.1132$ & $.5251$ & $.4610$ & $.5183$ & $.6372$ & $99.06\%$\\
  && Jaccard Sellers   							& $.\textbf{0158}$ & $.\textbf{0114}$ & $.\textbf{0154}$ & $.1092$ & $.1029$ & $.1047$ & $.5061$ & $.4940$ & $.4927$ & $.6054$ & $99.06\%$\\
    &&  Total Sellers   						& $.0065$ & $.0052$ & $.0073$ & $.0929$ & $.0743$ & $.0977$ & $.5079$ & $.4094$ & $.5113$ & $.6566$ & $99.06\%$ \\
    && Common Categories  					& $.0050$ & $.0039$ & $.0054$ & $.1090$ & $.1051$ & $.1041$ & $.5073$ & $.5366$ & $.4674$ & $.6123$ & $99.48\%$\\
  && Jaccard Categories   					& $.0058$ & $.0039$ & $.0049$ & $.\textbf{1361}$ & $.\textbf{1301}$ & $.\textbf{1288}$ & $.\textbf{5456*}$ & $.\textbf{5701*}$ & $.\textbf{5200*}$ & $.6364$ & $99.48\%$ \\
    &&  Total Categories   					& $.0007$ & $.0006$ & $.0009$ & $.0225$ & $.0236$ & $.0280$ & $.3317$ & $.3353$ & $.4215$ & $.6575$ & $99.48\%$ \\
    \hline 
  \multirow{12}{15pt}{\centering{\begin{sideways}\centering{Social}\end{sideways}}} &  
  \multirow{6}{15pt}{\centering{\begin{sideways}\centering{Content}\end{sideways}}} 
  &  Common Groups  								& $.0022$ & $.0010$ & $.0014$ & $.0402$ & $.0320$ & $.0425$ & $.3233$ & $.\textbf{2567}$ & $.3439$ & $.4307$ & $64.13\%$\\
  &&  Jaccard Groups 								& $.\textbf{0027}$ & $.\textbf{0016}$ & $.\textbf{0021}$ & $.\textbf{0433}$ & $.\textbf{0339}$ & $.\textbf{0459}$ & $.\textbf{3272}$ & $.2557$ & $.\textbf{3464}$ & $.4332$ & $64.13\%$\\
  &&  Total Groups   								& $.0006$ & $.0005$ & $.0007$ & $.0324$ & $.0272$ & $.0361$ & $.3214$ & $.2323$ & $.3563$ & $.4466$ & $64.13\%$\\
  &&  Common Interests  						& $.0005$ & $.0002$ & $.0002$ & $.0235$ & $.0201$ & $.0267$ & $.2285$ & $.1810$ & $.2474$ & $.3185$ & $46.51\%$\\
  &&  Jaccard Interests  						& $.0003$ & $.0002$ & $.0003$ & $.0219$ & $.0201$ & $.0227$ & $.2424$ & $.1857$ & $.2551$ & $.3161$ & $46.51\%$\\  
 &&  Total Interests   							& $.0004$ & $.0002$ & $.0003$ & $.0285$ & $.0235$ & $.0337$ & $.2639$ & $.1793$ & $.2765$ & $.3319$ & $46.51\%$\\
 \cline{2-14}
  &  \multirow{6}{15pt}{\centering{\begin{sideways}\centering{Network}\end{sideways}}} 
  &  Directed Interactions  				& $.0345$ & $.0389$ & $.0471$ & $.0582$ & $.0528$ & $.0630$ & $.1743$ & $.1756$ & $.1769$ & $.2169$ & $38.79\%$\\
  &&  Common Neighbors  						& $.1107$ & $.1021$ & $.1104$ & $.1216$ & $.1212$ & $.1243$ & $.2633$ & $.2887$ & $.2584$ & $.3300$ & $62.88\%$\\
  &&  Jaccard Common Neighbors 		& $.1381$ & $.1143$ & $.1378$ & $.1523$ & $.1386$ & $.1618$ & $.3726$ & $.3419$ & $.3814$ & $.4455$ & $71.53\%$\\
  &&  Neighborhood Overlap  				& $.\textbf{1434*}$ & $.\textbf{1222*}$ & $.\textbf{1471*}$ & $.\textbf{1620*}$ & $.\textbf{1486*}$ & $.\textbf{1719*}$ & $.3855$ & $.3471$ & $.3965$ & $.4514$ & $71.53\%$\\
  &&  Adamic/Adar  									& $.1013$ & $.0941$ & $.1067$ & $.1241$ & $.1187$ & $.1272$ & $.3028$ & $.3153$ & $.3042$ & $.3762$ & $69.34\%$\\
  &&  Pref. Attach. Score  					& $.0317$ & $.0331$ & $.0380$ & $.0630$ & $.0569$ & $.0650$ & $.3202$ & $.2838$ & $.3332$ & $.4420$ & $70.70\%$\\
  \hline 
  \multirow{14}{15pt}{\centering{\begin{sideways}\centering{Location}\end{sideways}}} &
  \multirow{9}{15pt}{\centering{\begin{sideways}\centering{Content}\end{sideways}}} 
  &  Common Favored Locations  			& $.0019$ & $.0010$ & $.0015$ & $.0427$ & $.0393$ & $.0481$ & $.4674$ & $.3773$ & $.4946$ & $.6437$ & $96.35\%$\\
  &&  Jaccard Favored Locations  		& $.0028$ & $.\textbf{0017}$ & $.0022$ & $.0472$ & $.\textbf{0416}$ & $.0531$ & $.4636$ & $.3777$ & $.4919$ & $.6490$ & $96.35\%$\\
    &&  Total Favored Locations			& $.\textbf{0031}$ & $.0016$ & $.\textbf{0023}$ & $.0459$ & $.0400$ & $.0513$ & $.4794$ & $.3802$ & $.5061$ & $.6635$ & $96.35\%$\\
  \cline{3-14}
  &&  Common Shared Locations   		& $.0003$ & $.0003$ & $.0004$ & $.0130$ & $.0103$ & $.0144$ & $.1449$ & $.1180$ & $.1599$ & $.2067$ & $30.45\%$\\
  && Jaccard Shared Locations  			& $.0005$ & $.0003$ & $.0004$ & $.0134$ & $.0115$ & $.0145$ & $.1420$ & $.1208$ & $.1522$ & $.2042$ & $30.45\%$\\
    &&  Total Shared Locations  		& $.0000$ & $.0000$ & $.0000$ & $.0092$ & $.0074$ & $.0106$ & $.1340$ & $.1076$ & $.1520$ & $.2031$ & $30.45\%$\\
\cline{3-14}  
  &&  Common Monitored Locations  	& $.0016$ & $.0010$ & $.0014$ & $.0408$ & $.0345$ & $.0449$ & $.4825$ & $.\textbf{3804}$ & $.5139$ & $.6734$ & $98.23\%$\\
  &&  Jaccard Monitored Locations   & $.0017$ & $.0008$ & $.0012$ & $.\textbf{0473}$ & $.0403$ & $.\textbf{0546}$ & $.\textbf{4987}$ & $.3795$ & $.\textbf{5387}$ & $.6760$ & $98.23\%$\\
    &&  Total Monitored Locations   & $.0011$ & $.0006$ & $.0009$ & $.0366$ & $.0331$ & $.0442$ & $.4770$ & $.3703$ & $.5354$ & $.6757$ & $98.23\%$ \\
  \cline{2-14}
  & \multirow{5}{15pt}{\centering{\begin{sideways}\centering{Network}\end{sideways}}} 
  &  Common Neighbors  						& $.0015$ & $.0007$ & $.0010$ & $.0298$ & $.0271$ & $.0345$ & $.3377$ & $.2623$ & $.\textbf{3632}$ & $.4609$ & $67.05\%$\\
  &&  Jaccard Common Neighbors			& $.\textbf{0016}$ & $.\textbf{0008}$ & $.\textbf{0011}$ & $.\textbf{0322}$ & $.0261$ & $.\textbf{0367}$ & $.3306$ & $.2623$ & $.3545$ & $.4579$ & $67.05\%$\\
  && Neighborhood Overlap   				& $.0014$ & $.0007$ & $.0010$ & $.0295$ & $.0267$ & $.0339$ & $.3359$ & $.2614$ & $.3608$ & $.4615$ & $67.05\%$\\
  &&  Adamic/Adar   								& $.0015$ & $.0006$ & $.0009$ & $.0320$ & $.\textbf{0272}$ & $.0353$ & $.3332$ & $.2598$ & $.3530$ & $.4595$ & $67.05\%$\\
  &&  Pref. Attach. Score   				& $.0003$ & $.0002$ & $.0002$ & $.0270$ & $.0213$ & $.0335$ & $.\textbf{3634}$ & $.\textbf{2825}$ & $.3458$ & $.4583$ & $70.59\%$\\
  \specialrule{.2em}{.1em}{.1em}
\end{tabular}
}
  \vspace{2mm}
\caption{Results for the user-based CF approaches based on various user similarity features showing their performance for the tasks of predicting products, low-level categories and top-level categories, respectively (\textit{RQ1}). \textit{Note:} Bold numbers indicate the highest accuracy values per feature set and ``*'' indicate the overall highest accuracy estimates.}
\vspace{4mm}
 \label{tab:msl_result_features_merged}
\end{center}
\end{sidewaystable*}

\begin{figure*}
  \centering
  		 \subfloat[Products]{ 
				\includegraphics[width=0.33\textwidth]{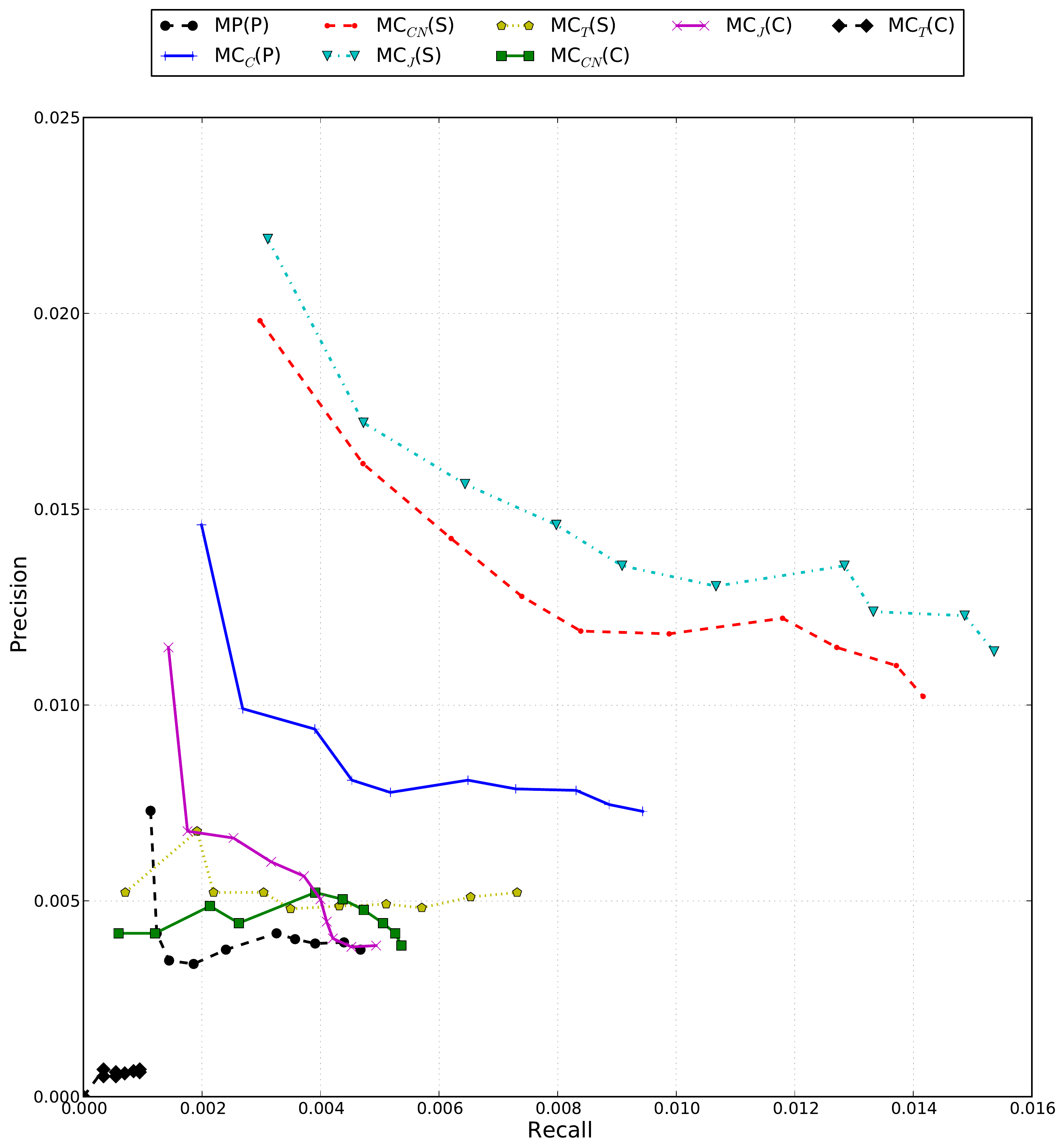}%
		 } 
		\subfloat[low-level categories]{ 
				\includegraphics[width=0.33\textwidth]{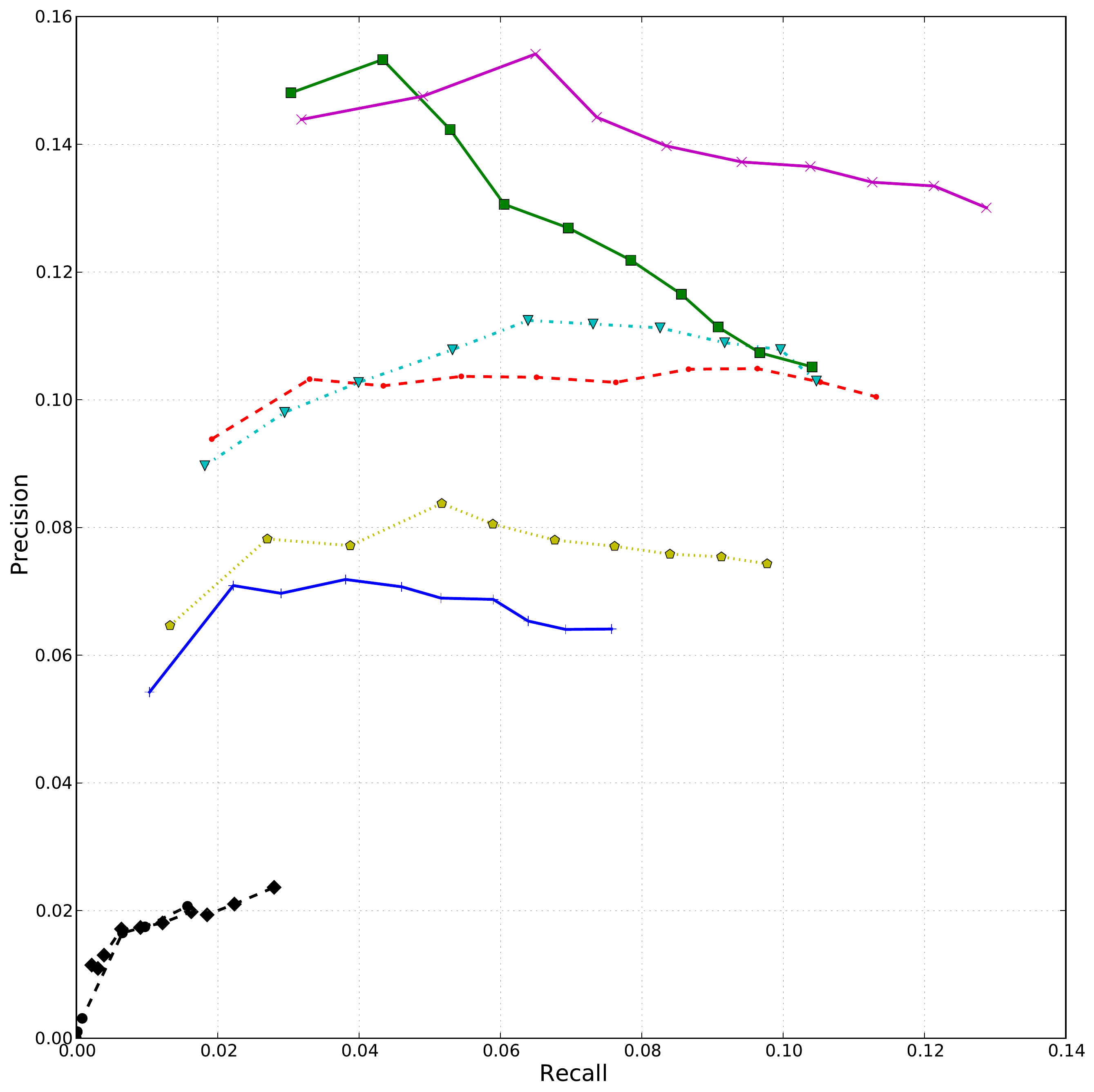}%
		 } 
		\subfloat[top-level categories]{ 
				\includegraphics[width=0.33\textwidth]{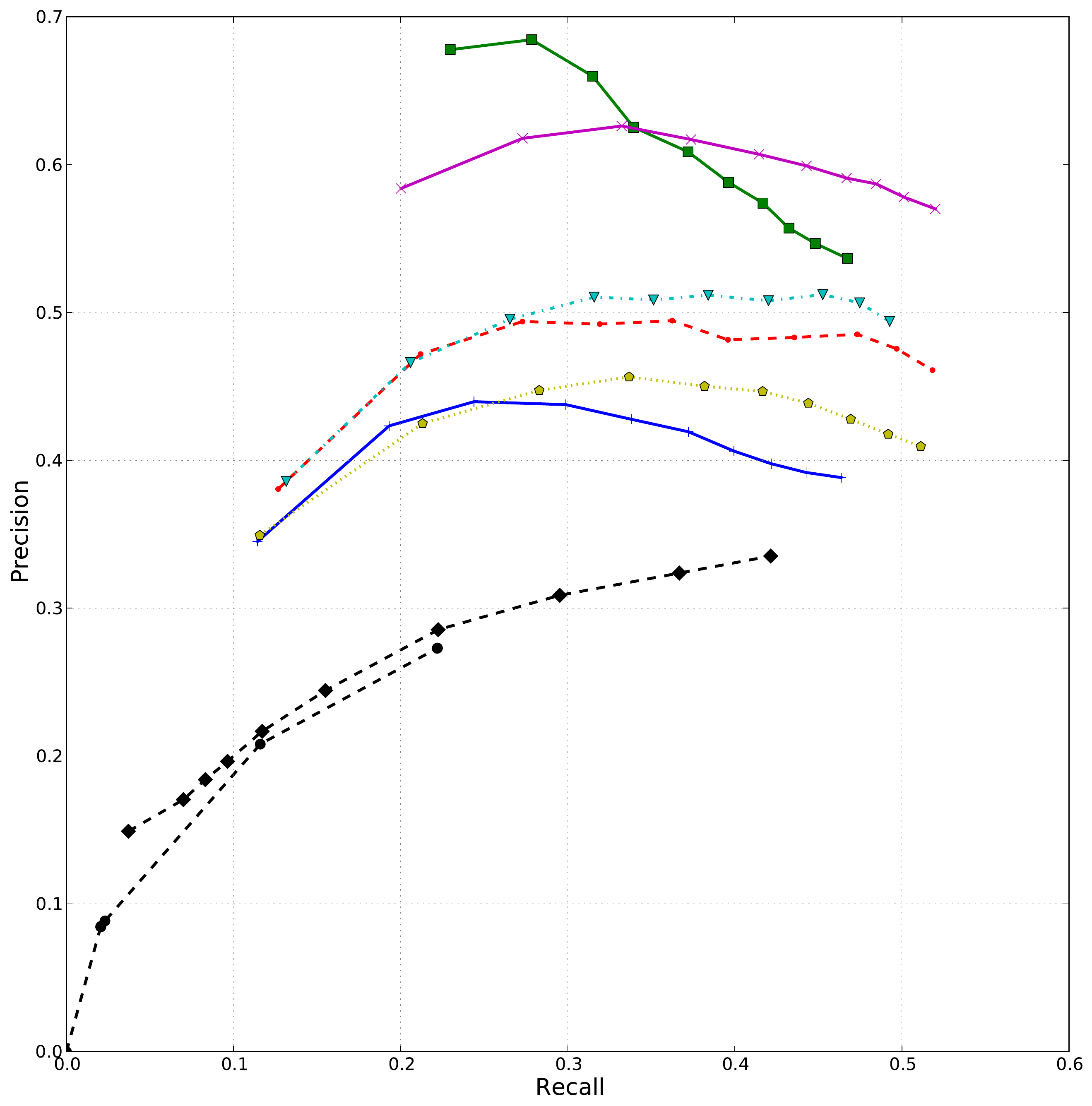}%
		 } 
		\\
		
		 \subfloat[Products]{ 
				\includegraphics[width=0.33\textwidth]{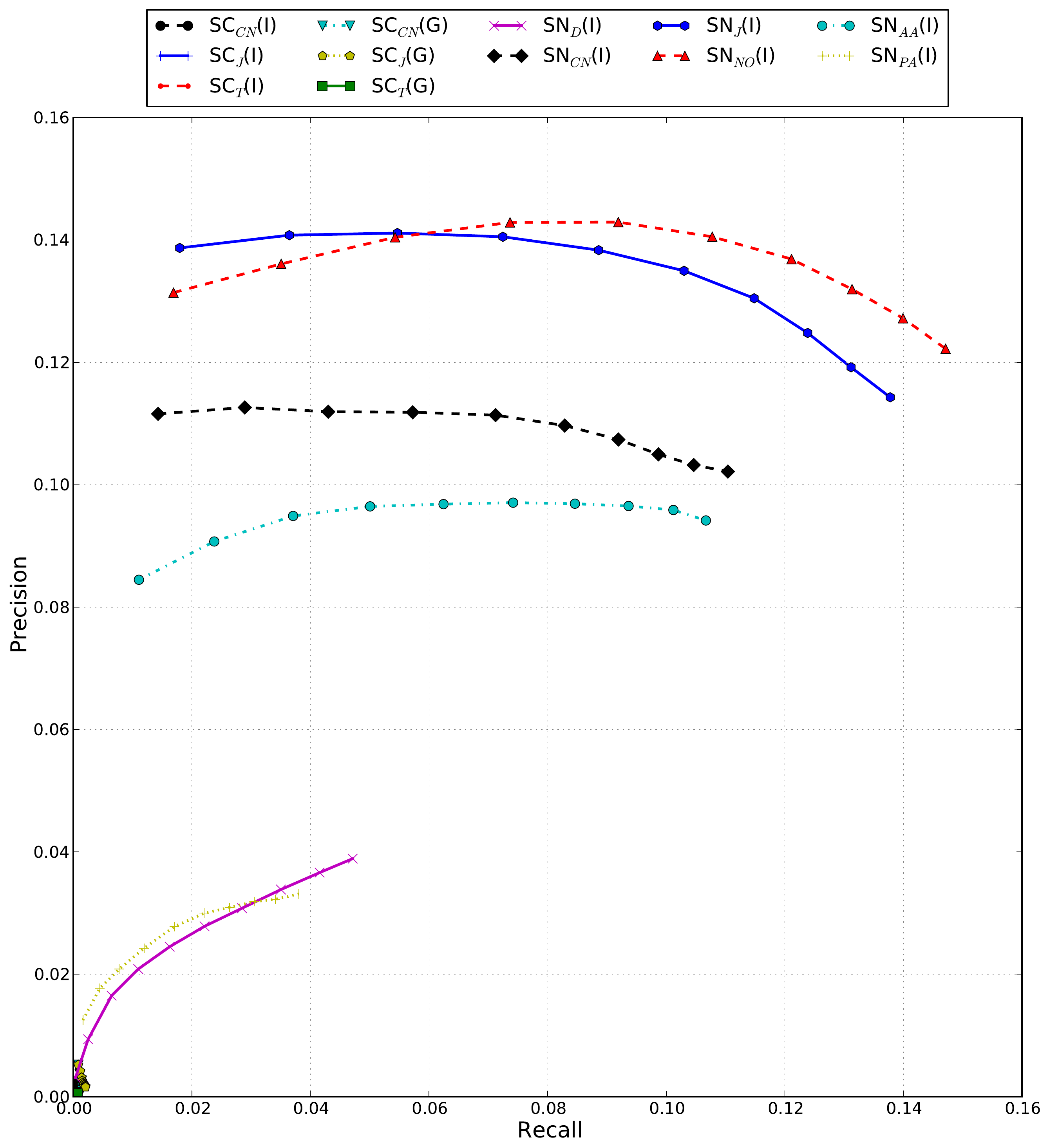}%
		 }
		\subfloat[low-level categories]{ 
				\includegraphics[width=0.33\textwidth]{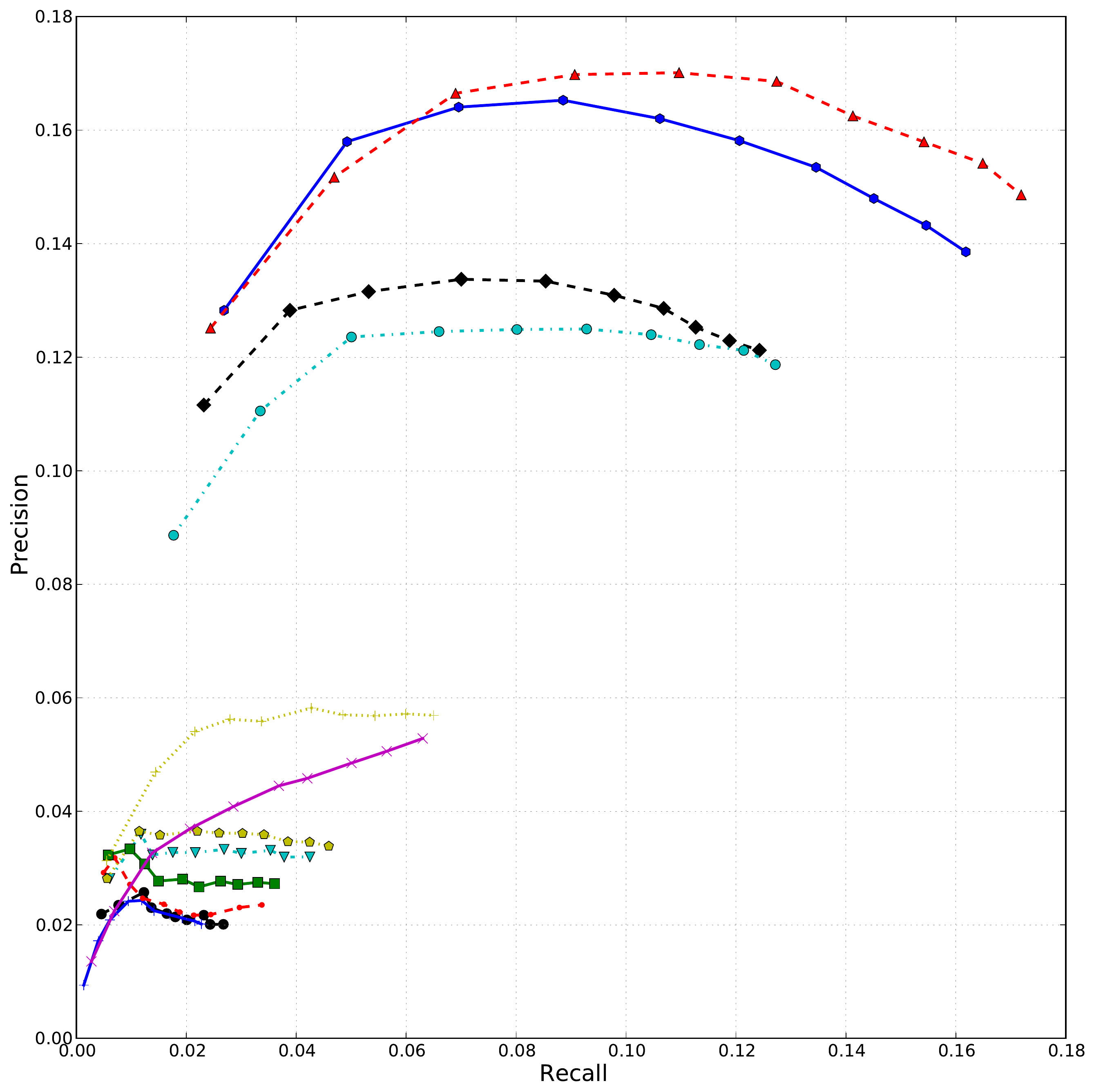}%
		 }
		\subfloat[top-level categories]{ 
				\includegraphics[width=0.33\textwidth]{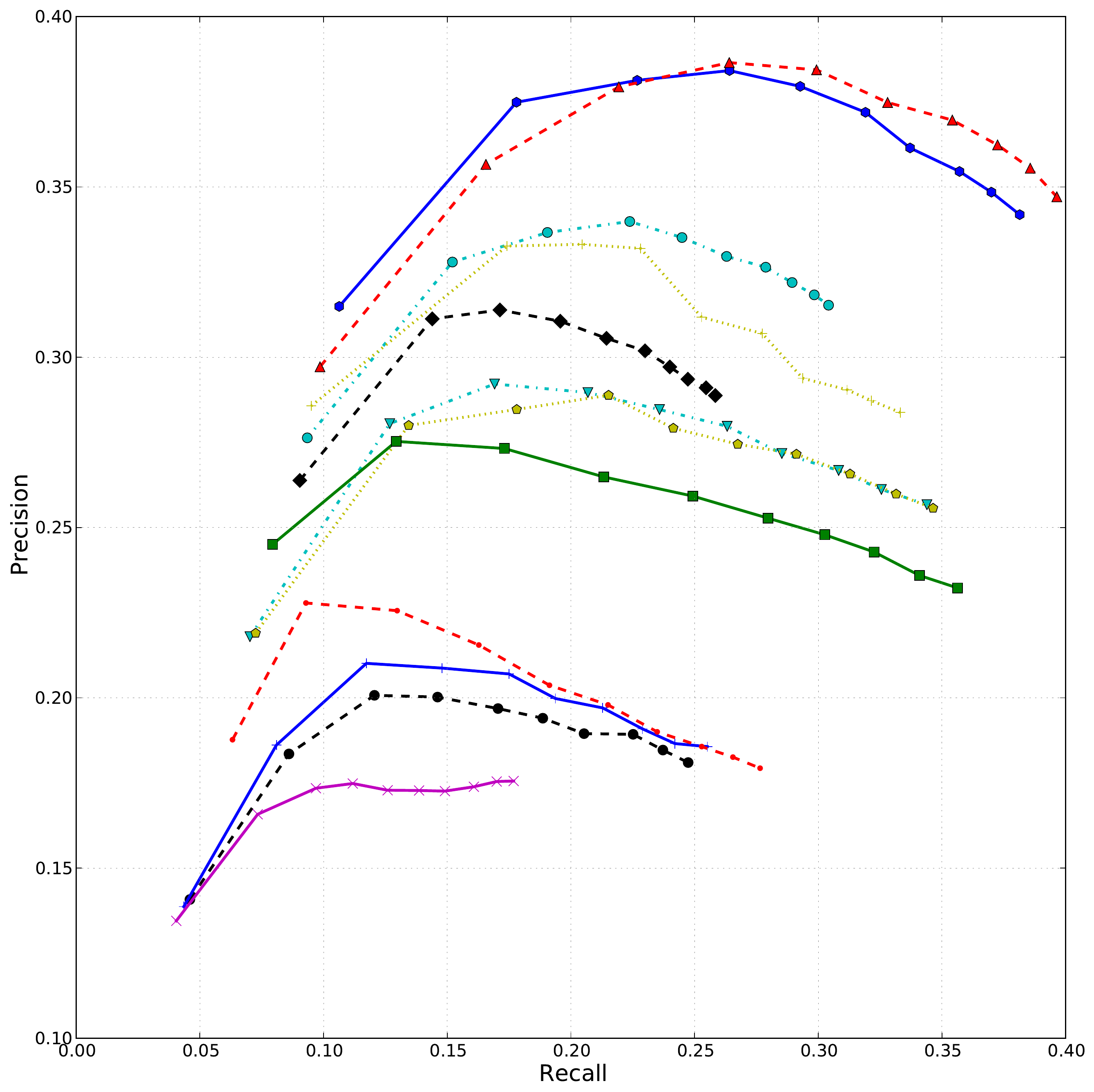}%
		 }
		\\
		 \subfloat[Products]{ 
				\includegraphics[width=0.33\textwidth]{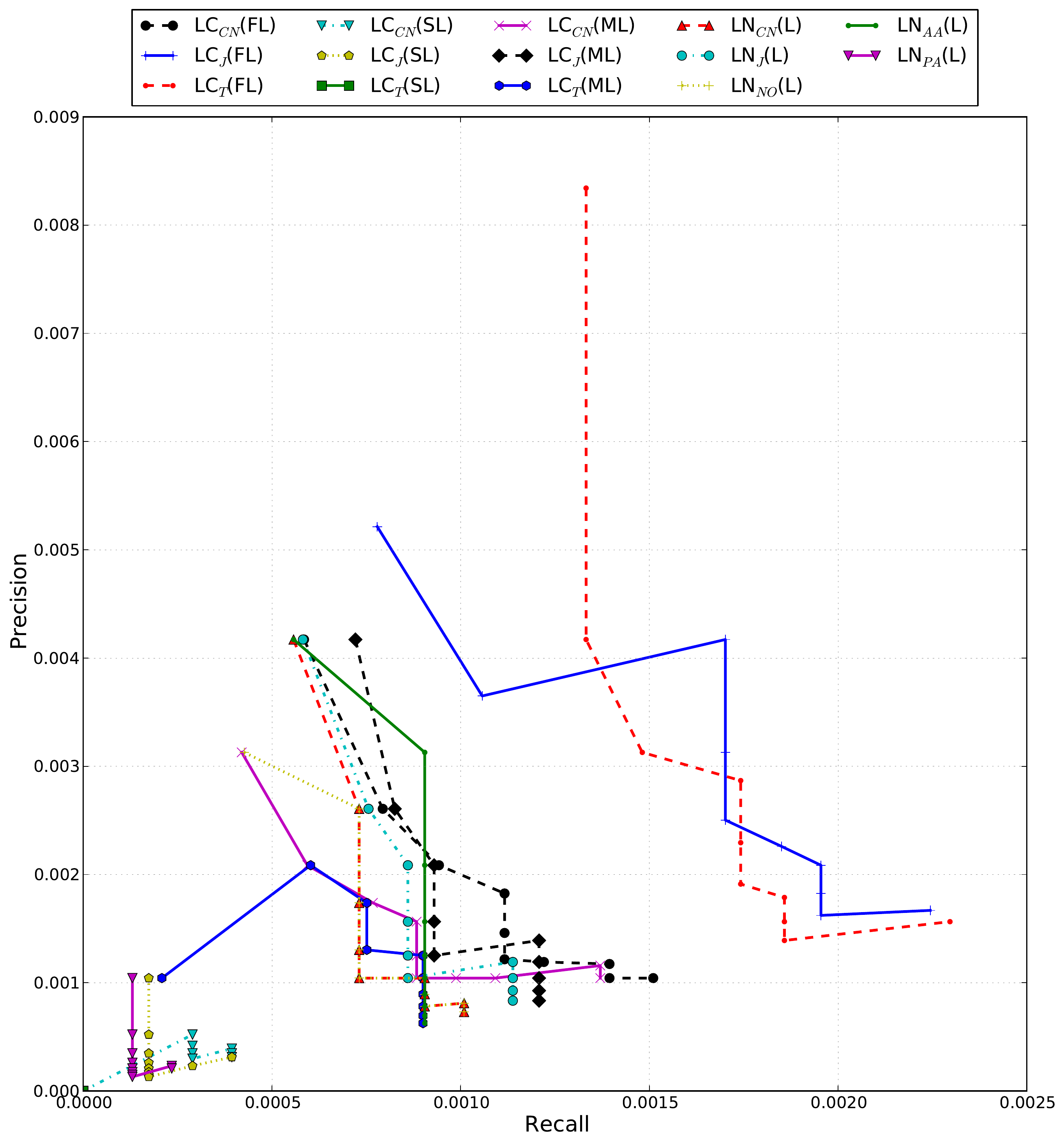}%
		 } 
		 \subfloat[low-level categories]{ 
				\includegraphics[width=0.33\textwidth]{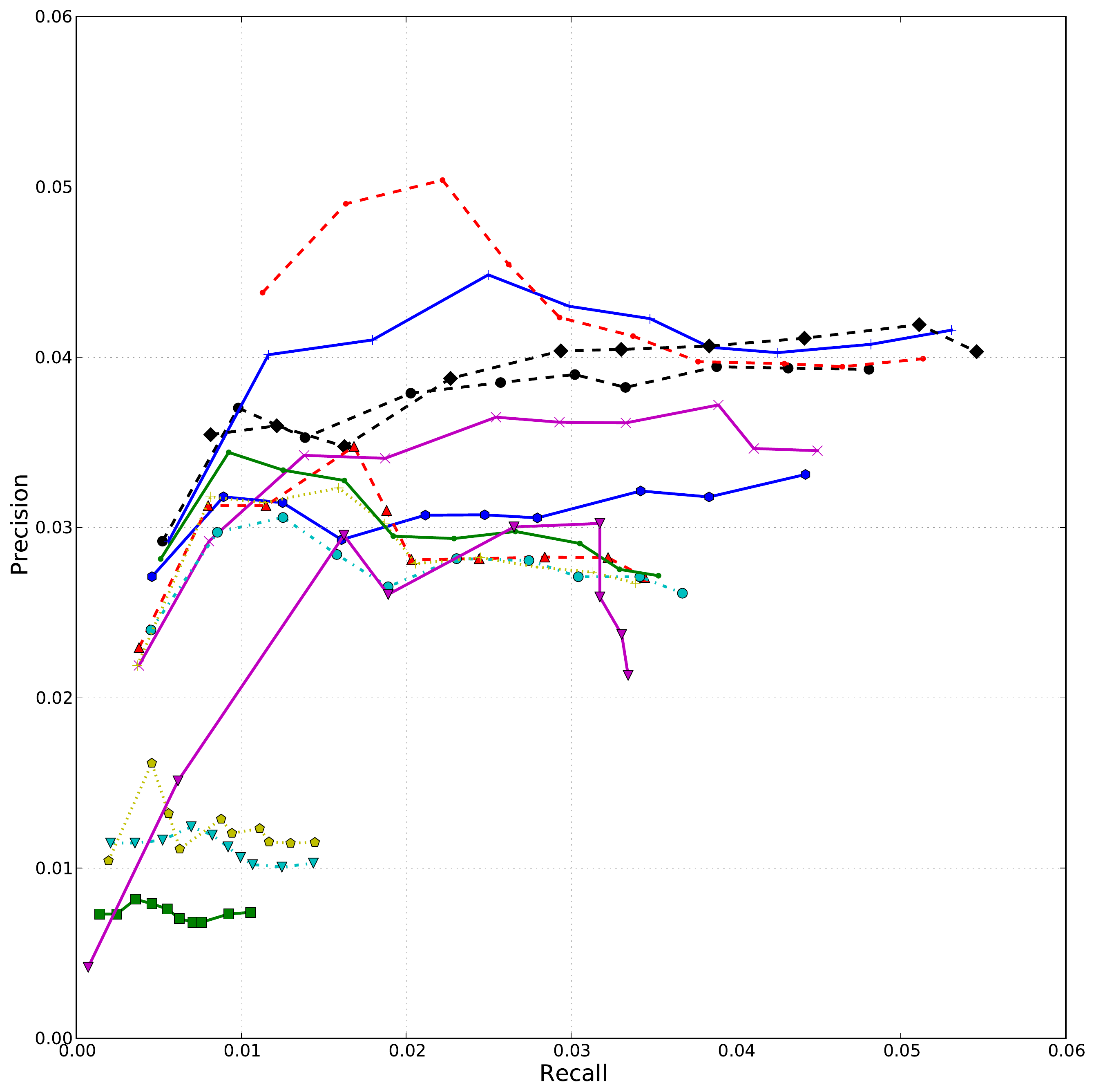}%
		 } 
		 \subfloat[top-level categories]{ 
				\includegraphics[width=0.33\textwidth]{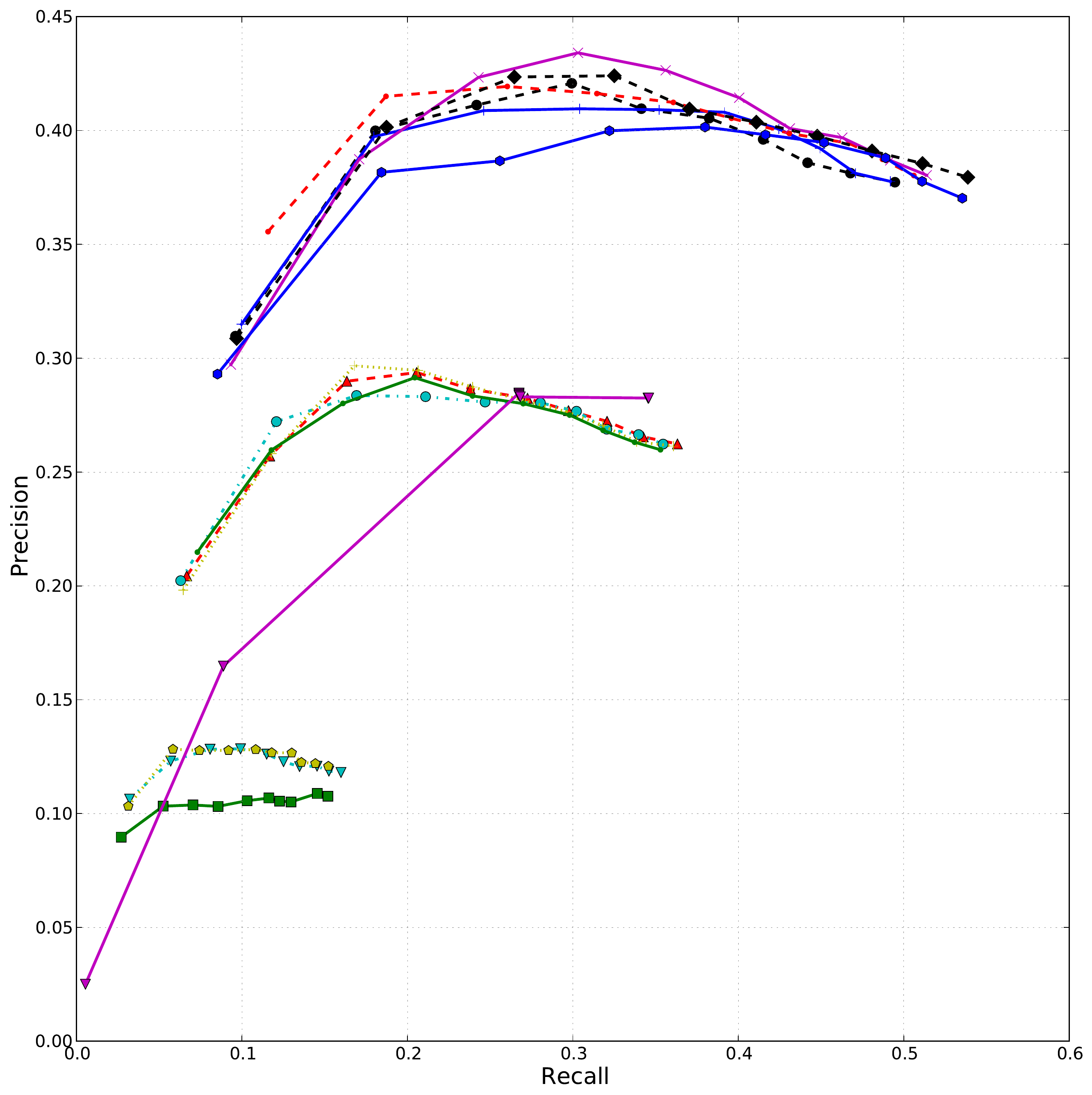}%
		 } 
     \caption{Recall/Precision plots for the single user similarity features derived from the marketplace (a, b, c), social network (d, e, f) and location-based (g, h, i) data sources, showing the performance of each feature for $k$ = 1 - 10 recommended items, low-level categories or top-level categories, respectively (\textit{RQ1}). \textit{Note:} The name for each feature in the legends is derived in the following way: the first two letters describe the data source, the subscript denotes the user similarity feature and the value in brackets defines the used data field (e.g., $SN_{NO}(I)$ stands for the Social Network data source, the Neighborhood Overlap similarity feature and Interactions data field).}
    \label{fig:recal_prec_features_normalized}
\end{figure*}

\section{Results}
\label{sec:eval}

In this section we highlight the results of our experiments for predicting products, low-level and top-level categories in terms of algorithmic performance in order to tackle our two research questions presented in Section \ref{sec:introduction}. Our evaluation has been conducted in two steps: in the first step we compared the different recommender approaches with the corresponding user similarity features isolated (\textit{RQ1}), see Table \ref{tab:msl_result_features_merged}) and in the second step we combined these approaches in the form of hybrid recommendations (\textit{RQ2}, see Table \ref{tbl:msl_result_hybrid_merged}). All results are presented by recommender accuracy, given by nDCG@10, P@10 (Precision) and R@10 (Recall), D@10 (Diversity) and UC (User Coverage). 

\subsection{Recommendations based on Single User Similarity Features}
The results for the recommendation of products, low-level categories and top-level categories using content-based and network-based user similarity features derived from our three data sources (marketplace, social and location-based data) are shown in Table \ref{tab:msl_result_features_merged} in order to address our first research question (\textit{RQ1}). Additionally, the results also include the \textit{Most Popular (MP)} approach as a baseline.

\textbf{Recommending Products} Regarding the task of predicting product purchases (first column in Table \ref{tab:msl_result_features_merged}), the best results in terms of recommender accuracy are reached by the network-based features based on interactions (e.g., loves, comments, wallposts) between the users in the social network. Surprisingly these approaches clearly outperform the user-based CF approaches relying on marketplace data, which implies that social interactions of the users are a better predictor to recommend products to people than marketplace data.


Another interesting finding is that the neighborhood-based features (\textit{Common Neighbors}, \textit{Jaccard}, \textit{Neighborhood Overlap} and \textit{Adamic/Adar}) also seem to be better indicators to determine the similarity between users than the direct interactions between these pairs. Only the \textit{Preferential Attachment Score} based recommender approach does not perform well in this context, although it still performs better than the features derived from the marketplace data source. This is to some extent expected and reveals that the individual's taste is more driven by the user's peers rather than by the popular users in the SecondLife social network.

In terms of the marketplace and location-based user similarity features, the results reveal that they do not provide high estimates of accuracy. This is interesting since our previous work \cite{steurer2013} showed that these features perform extremely well in predicting tie strength between users. However, the features derived from the marketplace and the location-based data sources provide the best results with respect to Diversity (D) and User Coverage (UC). 

\textbf{Recommending categories.} Regarding the tasks of predicting low-level and top-level categories, the second and third column of Table \ref{tab:msl_result_features_merged} report the accuracy estimates for the different user similarity features based on the extracted categories. As expected, all user similarity features end up with a much higher accuracy than for predicting products, especially in the case of top-level categories, because of the lower level of specialization of these recommendation tasks. In the case of the low-level category predictions, the approaches based on social interaction features still perform better than the approaches based on features of the marketplace or location-based data sources. Interestingly, the content-based user similarity features derived from the social network as well as the location-based features, which performed the worst at product predictions, perform much better for low-level categories, now also outperforming the \textit{MP} baseline.

In the case of the top-level category recommendations, it can be seen that the user similarity features of all three data sources provide quite similar results in terms of recommender accuracy. The approach based on the Jaccard's coefficient for categories performs best in terms of nDCG@10, P@10 and R@10. This result is very interesting since this feature is based on the marketplace data source that provided quite bad results in the case of product predictions. Summed up, we see that user similarity features derived from all data sources are very useful indicators for recommendations, although they depend on the level of specialization of the recommendation task (\textit{RQ1}).

\subsection{Recommendations Based on Combined Data Sources}
\begin{figure*}[t!]
		\centering
  		 \subfloat[Products]{ 
				\includegraphics[width=0.33\textwidth]{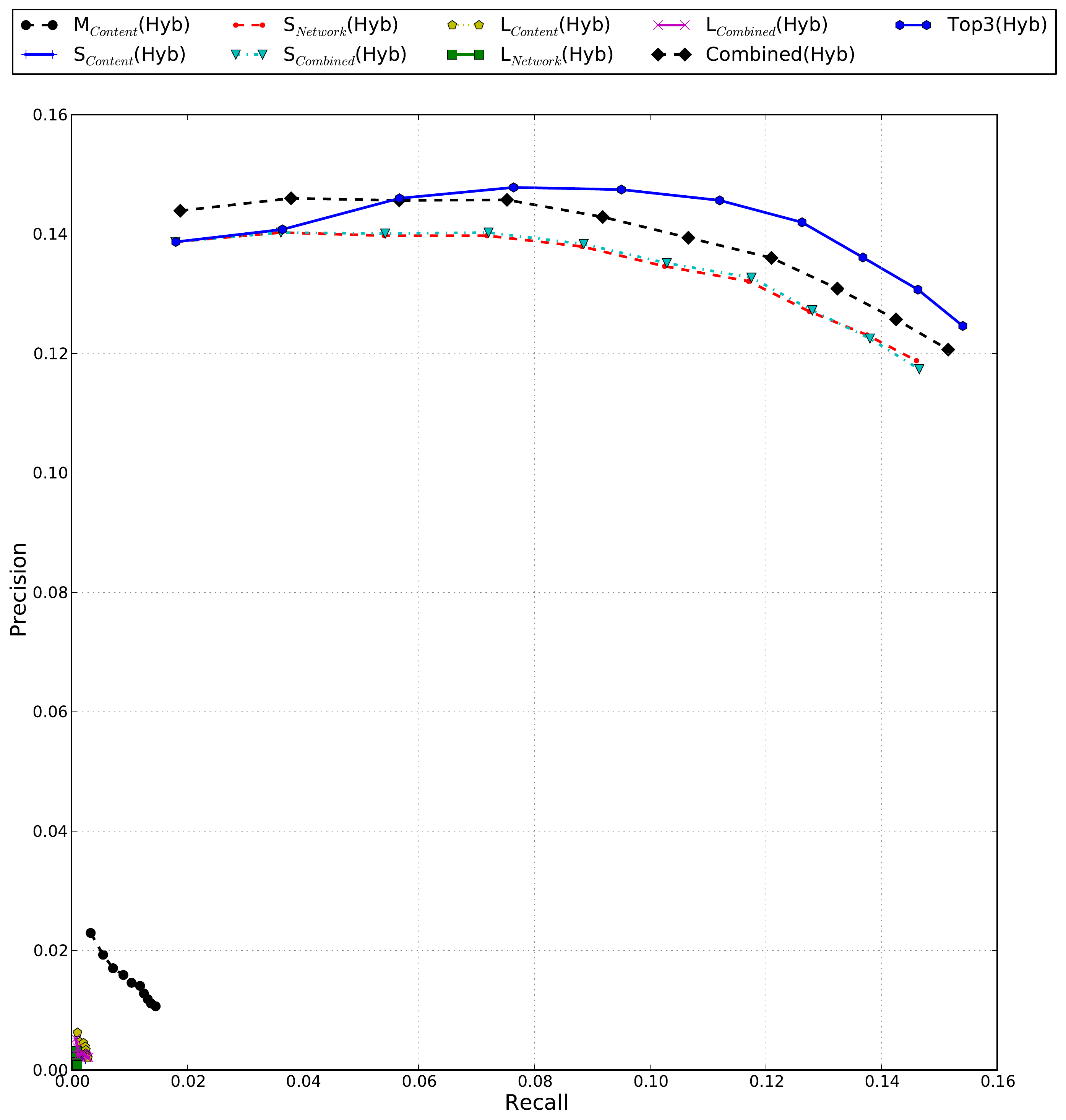}%
		 } 
		 \subfloat[low-level categories]{ 
				\includegraphics[width=0.33\textwidth]{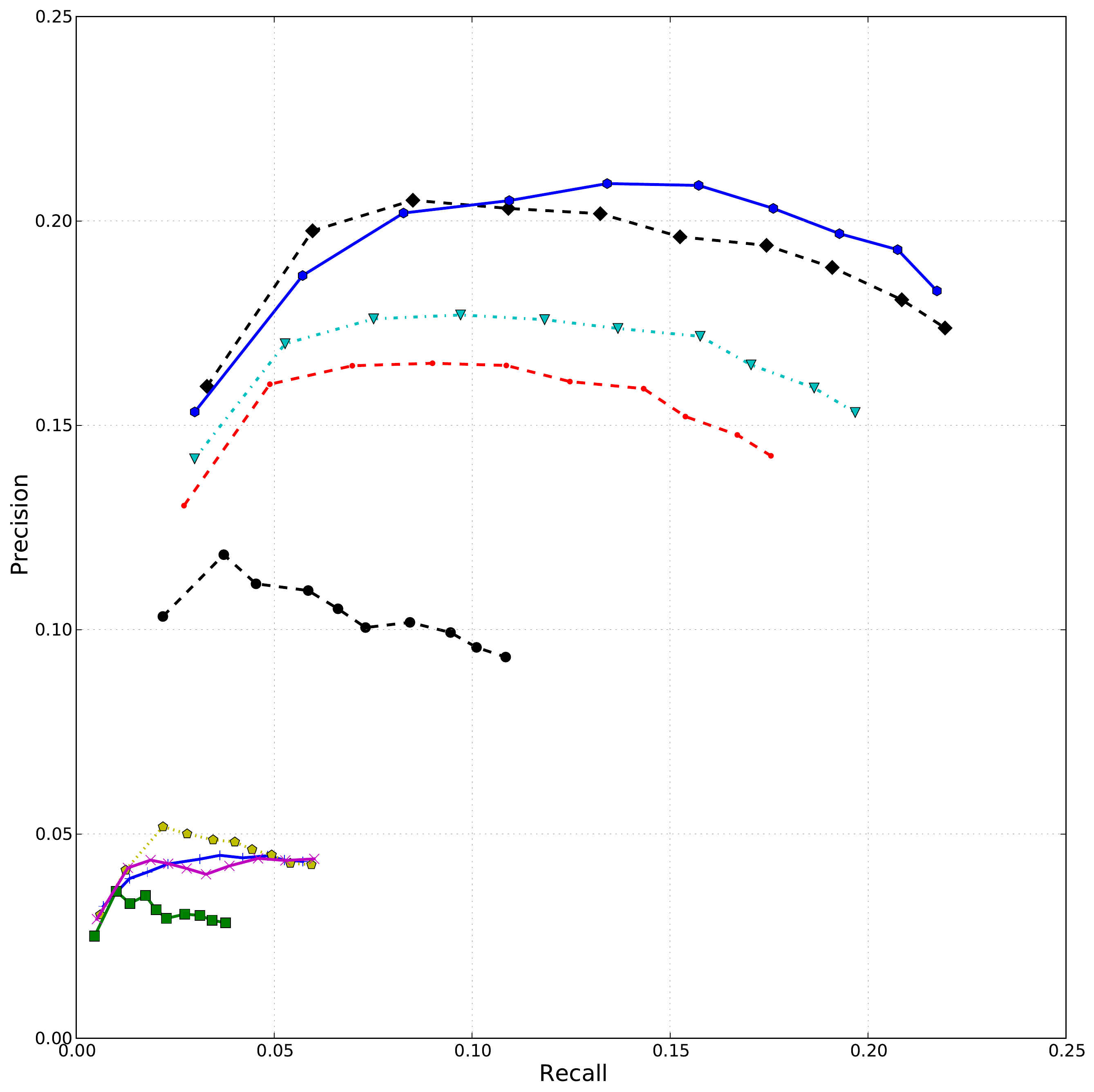}%
		 }  
		 \subfloat[top-level categories]{ 
				\includegraphics[width=0.33\textwidth]{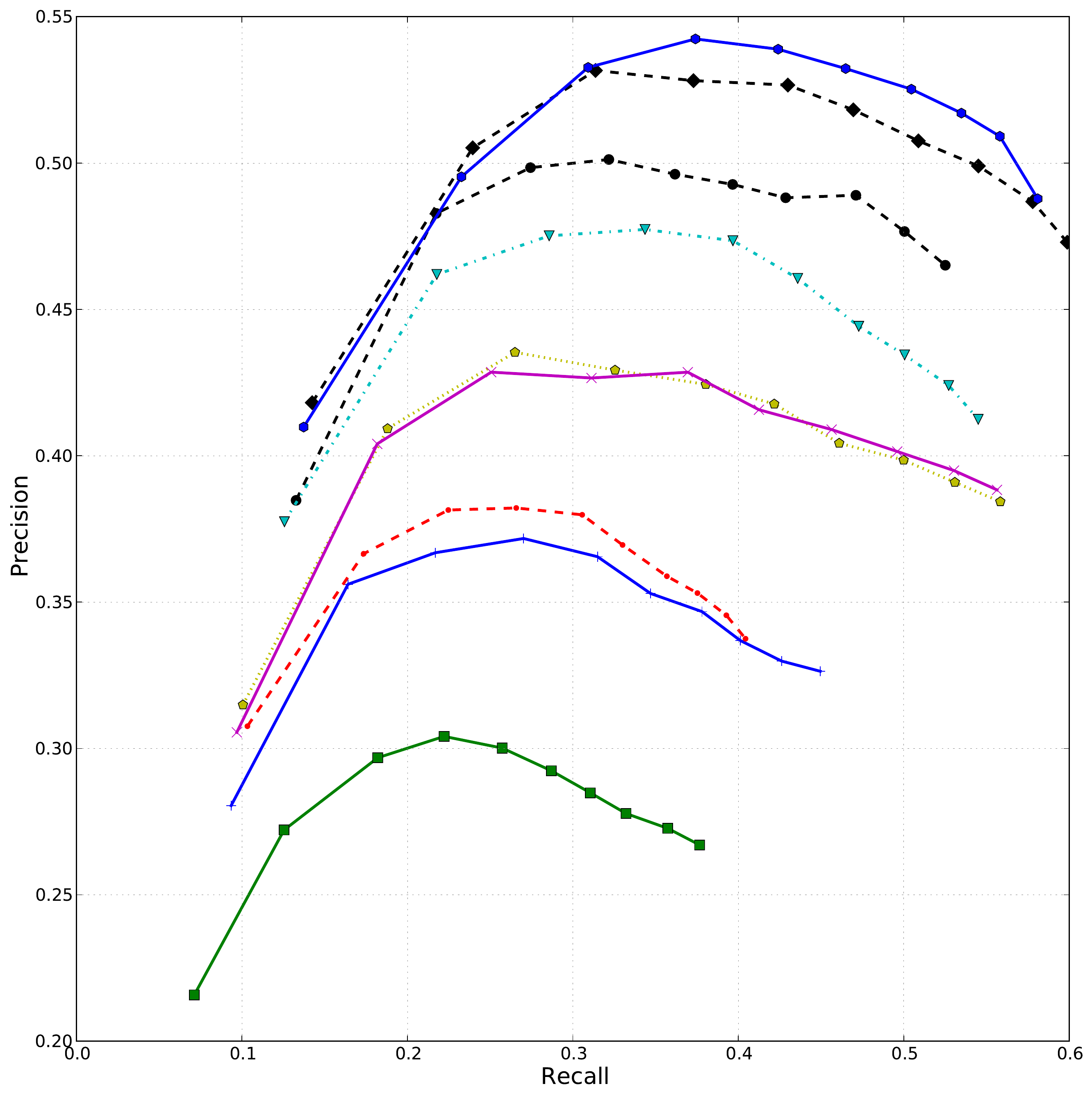}%
		 }
     \caption{Recall/Precision plots for the hybrid approaches showing the performance of each data source for $k$ = 1 - 10 recommended items, low-level categories or top-level categories, respectively (\textit{RQ2}).}
    \label{fig:recal_prec_hybrid}
\end{figure*}

The findings of the last subsection suggests that a combination of features from all three data sources (marketplace, social network and location-based data) should provide more robust recommendations in case of both tasks, predicting products and categories (\textit{RQ2}). Thus, Table \ref{tbl:msl_result_hybrid_merged} shows the results of the hybrid approaches based on theses data sources in order to tackle our second research question. As before, the first column indicates the results for the product prediction, the second column for the low-level category prediction and the third column for the top-level category prediction. Additionally, Figure \ref{fig:recal_prec_hybrid} shows the performance of the hybrid approaches based on the three data sources for $k$ = 1 - 10 recommended items, low-level categories or top-level categories, respectively, in form of Recall/Precision plots.

\begin{sidewaystable*}
\hspace*{-1mm}
\vspace*{0mm}
  \begin{center}
\scalebox{0.85}{
  \setlength{\tabcolsep}{6pt}
    \begin{tabular}{l|l|l|ccc|ccc|ccc|cc} 
    \specialrule{.2em}{.1em}{.1em}
				\multicolumn{3}{c}{\centering{}}  & \multicolumn{3}{c}{\centering{Products}} & \multicolumn{3}{c}{\centering{low-level categories}} & \multicolumn{3}{c}{\centering{top-level Categories}} & & \\ \hline

   & \multicolumn{2}{c|}{Sets}& $nDCG@10$ 	   & $P@10$ 		& $R@10$ & $nDCG@10$ 	   & $P@10$ 		& $R@10$ & $nDCG@10$ 	   & $P@10$ 		& $R@10$ & $D@10$  & $UC$\\ \hline

    & \multicolumn{2}{c|}{ Most Popular }  	  						& $.0082$ & $.0021$	 &	$.0122$ & $.0185$ & $.0207$ & $.0157$ & $.2380$ & $.2730$ & $.2221$ & $.5945$  & $100.00\%$\\

  \hline\hline
  \multirow{9}{*}{\centering{\begin{sideways}\centering{Weighted Sum}\end{sideways}}} &
        \multirow{1}{*}{\parbox{1.3cm}{\centering{Market}}}
                      & Content 							& $.0158$ & $.0107$ & $.0145$ & $.1109$ & $.0933$ & $.1085$ & $.\textbf{5292}$ & $.\textbf{4651}$ & $.5251$ & $.6389$ & $99.79\%$ \\ \cline{2-14} 
    & \multirow{3}{*}{\parbox{1.3cm}{\centering{Social}}}
                      & Content 							& $.0029$ & $.0018$ & $.0024$ & $.0527$ & $.0433$ & $.0572$ & $.4216$ & $.3264$ & $.4495$  & $.5589$ & $81.65\%$ \\         
                    &  & Network 							& $.1422$ & $.\textbf{1188}$ & $.1460$ & $.1643$ & $.1425$ & $.1755$ & $.3940$ & $.3375$ & $.4044$  & $.4589$ & $71.53\%$ \\             
                   &  & Combined    					& $.\textbf{1427}$ & $.1174$ & $.\textbf{1465}$ & $.\textbf{1821}$ & $.\textbf{1531}$ & $.\textbf{1968}$ & $.5223$ & $.4125$ & $.5450$  & $.6241$ & $92.70\%$\\ \cline{2-14}   
    & \multirow{3}{*}{\parbox{1.3cm}{\centering{Location}}}
                     & Content 								& $.0035$ & $.0020$ & $.0028$ & $.0531$ & $.0425$ & $.0594$ & $.5139$ & $.3843$ & $.\textbf{5583}$  & $.6963$ & $100.00\%$ \\       
                    &  & Network  						& $.0015$ & $.0007$ & $.0010$ & $.0330$ & $.0283$ & $.0377$ & $.3535$ & $.2670$ & $.3766$ & $.4864$ & $70.59\%$\\     
                    &  & Combined   					& $.0031$ & $.0021$ & $.0030$ & $.0537$ & $.0439$ & $.0601$ & $.5152$ & $.3883$ & $.5563$  & $.6932$ & $100.00\%$\\ \cline{2-14} 
   & \multicolumn{2}{c|}{Combined}  					& $.1477$ & $.1206$ & $.1515$ & $.\textbf{2086*}$ & $.1740$ & $.\textbf{2197*}$ & $.\textbf{5868*}$ & $.4734$ & $.\textbf{5994*}$ & $.6630$ & $100.00\%$\\
    &\multicolumn{2}{c|}{Combined Top 3}  		& $.\textbf{1498*}$ & $.\textbf{1246*}$ & $.\textbf{1540*}$ & $.2078$ & $.\textbf{1829*}$ & $.2174$ & $.5696$ & $.\textbf{4878*}$ & $.5809$  & $.6521$ & $100.00\%$\\
 
    \specialrule{.2em}{.1em}{.1em}

  \end{tabular}
  }
  \end{center}
   \vspace{2mm}
  \caption{Results for the hybrid approaches based on our three data sources for the tasks of predicting products, low-level categories and top-level categories. The results show that all three data sources (marketplace, social network and location-based data) are important indicators for calculating recommendations, since a hybrid combination of all data sources provided the best results in case of predicting products, low-level categories and top-level categories (\textit{RQ2}). \textit{Note:} Bold numbers indicate the highest accuracy values across the data sources and ``*'' indicate the overall highest accuracy estimates.}
 \vspace{4mm}  
\label{tbl:msl_result_hybrid_merged}
\end{sidewaystable*}

\textbf{Recommending products.} Regarding the product prediction task, we see again that the recommender approaches based on the social network data source clearly outperform the ones based on the marketplace and location-based data sources as well as the \textit{MP} baseline. Furthermore, when combining all three data sources, not only the overall recommendation accuracy is increased with respect to nDCG@10, P@10 and R@10, but also the User Coverage (UC) is increased to the maximum of 100\%.

This means that the hybrid approach combines the strengths of the user similarity features of all three data sources in order to be capable of providing accurate recommendations for all users in the datasets. Another hybrid approach shown in Table \ref{tbl:msl_result_hybrid_merged} combines only the best user similarity features from each data source (referred to as \textit{Top 3}) and reaches higher accuracy estimates, but a smaller Diversity (D), than the hybrid approach that combines all user similarity features.

\textbf{Recommending categories.} In contrast to the results of the product predictions, that showed that the recommender based on the social network data source clearly outperforms the recommenders based on the marketplace and location-based data sources, the results of the category predictions (second and third column of Table \ref{tbl:msl_result_hybrid_merged} do not show that big differences between the three data sources. With respect to the low-level category predictions, we again observe that the recommender based on the social network data source still provides the highest accuracy estimates.

Interestingly, in this case also the recommenders based on the other two data sources provide reasonable results, which has not been the case of predicting products, where the location-based recommender even was outperformed by the \textit{MP} baseline. Based on these results we would assume that marketplace and location-based data sources are suitable of providing accurate predictions in more general recommendation tasks. The results for the top-level category predictions prove this assumption since in this case the recommenders based on marketplace and location-based data sources even provide better results in terms of recommender accuracy, Diversity and User Coverage than the one based on social network data in case of the content-based features. As before, the combination of all three data sources provide again the best results. Summed up, this results prove our assumption derived from \textit{RQ1}, that all three data sources are important for calculating recommendations, since a combination of all data sources provided the best results in case of predicting products, low-level categories and top-level categories (\textit{RQ2}).

%
%

\section{Conclusions \& Future Work} \label{sec:con}
In this work we presented first results of a recently started project that tries to utilize various user similarity features derived from three data sources (marketplace, social network and location-based data) to recommend products and points of interests (i.e., low-level and top-level categories) to people in an online marketplace setting. This section concludes the paper with respect to our two research questions and gives an outlook into the future.

The first research question of this work (\textit{RQ1}) dealt with the question as to which extent user similarity features derived from marketplace, social network and location-based data sources can be utilized for the recommendation of products and categories in online marketplaces. To tackle this question we implemented various user-based Collaborative Filtering (CF) approaches based on the user similarity features from the data sources and tested them isolated. As the results have shown, the user-based CF approaches that utilize features of online social network data to calculate the similarities between users performed best in case of predicting products, significantly outperforming the other approaches relying on both -- marketplace and location-based user data. However, this behavior changed in the case of predicting low-level and top-level categories where the differences between the three data sources got substantially smaller. Surprisingly, with respect to the top-level category predictions, the marketplace and location-based features even reached the highest results in terms of accuracy, Diversity (D) and User Coverage (UC).

These results showed that user similarity features of all three data sources are important indicators for recommendations and suggests that combining them should result into more robust recommendations, especially in cases of multiple recommendation tasks on different levels of specialization (topics and categories). Thus, our second research question (\textit{RQ2}) tried to tackle the question if the different marketplace, social network and location-based user similarity features and data sources can be combined in order to create a hybrid recommender that provides more robust recommendations in terms of prediction accuracy, diversity and user coverage. In order to address this question we implemented and evaluated hybrid recommenders that combined the features of the data sources. The results proved our assumption and showed that hybrid recommender that combined user similarity features of all three data sources provided the best results across all accuracy metrics (nDCG@10, P@10, R@10) and all settings (product, low-level category and top-level category recommendations). Moreover, this hybrid recommender also provided a User Coverage of 100\% and thus, is able to provide these accurate recommendation to all users in the datasets.

Although the results of this study are based on a dataset obtained from the virtual world SecondLife, we believe that it bears great potential to create a sequence of interesting studies that may have implications for the ``real'' world (see e.g., \cite{szell2012understanding}). For instance, one of the potential interesting issues we are currently exploring is predicting products and categories to users in a cold-start setting (i.e., for users that only have purchases a few or even no products in the past) by a diversity of features. Other important work we plan is the use of state-of-the-art model-based approaches in order to assess whether the signals extracted from similarity features in the current analysis can be replicated (in the case of social data) or improved (in the case of location data) for different recommendation tasks.

We have also shown that using the interaction information between users improves not only the task of product recommendation, but also the recommendation of low-level and top-level categories. Thus, we are also interested in studying the extent to which recommendations can be improved by utilizing content-based similarity features derived from the users' social streams.

\subsubsection{Acknowledgments}
This work is supported by the Know-Center and the EU funded project Learning Layers (Grant Agreement 318209). 
The Learning Layers project is supported by the European Commission within the 7th Framework Program, under the DG Information society and Media (E3), unit of Cultural heritage and technology-enhanced learning.
The Know-Center is funded within the Austrian COMET Program - Competence Centers for Excellent Technologies - under the auspices of the Austrian Ministry of Transport, Innovation and Technology, the Austrian Ministry of Economics and Labor and by the State of Styria. COMET is managed by the Austrian Research Promotion Agency (FFG).

\bibliographystyle{splncs}
\bibliography{sbc}

\end{document}